\documentclass[preprint]{elsarticle}

\usepackage{graphicx}
\usepackage{amssymb}
\usepackage{tabularx}
\usepackage{xcolor}
\journal{Journal of Solid State Chemistry}

\begin{document}

\begin{frontmatter}

\title{Structural transitions and spontaneous exchange bias in La$_{2-x}$Ba$_{x}$CoMnO$_{6}$ series}

\author[AA]{H. Fabrelli}, \author[BB]{A. G. Silva}, \author[BB]{M. Boldrin}, \author[BB]{L. Bufai\c{c}al\corref{Bufaical}}, \ead{lbufaical@ufg.br}, \author[AA]{E. M. Bittar}

\address[AA]{Centro Brasileiro de Pesquisas F\'{\i}sicas, Rua Dr. Xavier Sigaud 150, 22290-180, Rio de Janeiro, RJ, Brazil}
\address[BB]{Instituto de F\'{i}sica, Universidade Federal de Goi\'{a}s, 74001-970 , Goi\^{a}nia, GO, Brazil}

\cortext[Bufaical]{Corresponding author}

\begin{abstract}
The structural, electronic, and magnetic properties of La$_{2-x}$Ba$_{x}$CoMnO$_{6}$ ($0.25 \leq x \leq 1.0$) compounds were investigated employing x-ray powder diffraction and magnetometry. The crystal structure evolves from orthorhombic $Pnma$ to rhombohedral $R\bar{3}c$ and then to hexagonal $P6_{3}/mmc$ space group as the Ba concentration increases. The magnetization as a function of temperature measurements revealed a ferromagnetic-like behavior for all samples, but antiferromagnetic interactions seem to also be present throughout the whole series. The competition between magnetic phases leads to the phenomenon of spontaneous exchange bias effect, observed in the zero field cooled magnetization as a function of applied magnetic field curves. The evolution of this effect along the series is discussed in terms of changes in the crystal structure.
\end{abstract}


\end{frontmatter}

\section{INTRODUCTION}

The discovery of the spontaneous exchange bias (SEB) effect \cite{Saha,Wang}, \textit{i.e.} a shift in the magnetization as a function of applied magnetic field [M(H)] curve observed even when the system is cooled in the absence of external field, has opened the path for novel applications of the EB phenomena, once the unidirectional magnetic anisotropy can be set without the assistance of a cooling field. The presence of a spin-glass (SG)-like phase concomitant to other conventional magnetic phases seems to be a necessary condition for the emergence of such effect \cite{Model,Model2,PRM2021}. In this regard, double-perovskite (DP) compounds are auspicious candidates for discovering and investigating new SEB materials. This interest in DP materials is because they present two transition-metal (TM) ions in their structure, usually leading to disorder and competition between magnetic phases, the ingredients for the emergence of glassy magnetism \cite{Mydosh}.

Among the DP compounds that present the SEB effect, the CoMn-based compounds present the most robust magnetization shifts \cite{Pal,Giri,APL2020,Zhang}. The La$_{2-x}$Sr$_{x}$CoMnO$_{6}$ series was the first discovered SEB DP system, for which the evolution of the exchange bias field ($H_{EB}$) along the series was attributed to changes in the anti-site disorder (ASD) at Co/Mn site \cite{Murthy,Murthy2}. Later on, La$_{2-x}$Ca$_{x}$CoMnO$_{6}$ was also shown to exhibit SEB, although with a much smaller magnitude than that of Sr-based series \cite{JMMM2017}. Such reduction in the SEB effect is believed to be related to the strong crystal field that inhibits the high spin (HS) configuration for Co$^{3+}$ in the Ca-based system \cite{PRB2019}. For the latter series, the similarity between the ionic radii of Ca$^{2+}$ and La$^{3+}$ \cite{Shannon} leads to less pronounced changes in the crystal structure with hole doping. The systematic increase of the SEB effect with Ca-doping is attributed to the strengthening of the uncompensated magnetic coupling at the magnetic interfaces caused by the increased antiferromagnetic (AFM) phase \cite{LaCa2022}.

Contrastingly, doping the La$^{3+}$ with Ba$^{2+}$ instead of Ca$^{2+}$ is expected to impact significantly not only the electronic of the TM ions but also the crystal structure. Here we investigate how the structural and electronic changes caused by such substitution affect the magnetic properties of La$_{2-x}$Ba$_{x}$CoMnO$_{6}$ ($0.25 \leq x \leq 1.0$) series. Employing x-ray powder diffraction (XRD), we observe two structural transitions along the series, with the crystal structure evolving from orthorhombic $Pnma$ to rhombohedral $R\bar{3}c$, and then to hexagonal $P6_{3}/mmc$ space group, as the Ba-content increases. DC magnetization as a function of temperature [M(T)] measurements confirmed the ferromagnetic (FM) ordering for all investigated samples. However, the presence of Co and Mn valence changes due to the hole doping leads to the emergence of AFM interactions. The competing magnetic interactions give rise to the SEB effect observed in the M(H) curves. The evolution of this effect seems to be directly related to the system's crystal symmetry.

\section{EXPERIMENT DETAILS}

Polycrystalline samples of La$_{2-x}$Ba$_{x}$CoMnO$_{6}$ series ($x$ = 0.25, 0.5, 0.75, 1.0) were synthesized by conventional solid-state reaction method. Stoichiometric amounts of La$_{2}$O$_{3}$, BaCO$_3$, Co$_{3}$O$_{4}$ and MnO were mixed and heated at $700^{\circ}$C for 10 hours in air atmosphere, followed by a second treatment at $1000^{\circ}$C for 20 hours. Later, the samples were re-grinded before a step of 60 hours at $1200^{\circ}$C, with intermediate grinding. Finally, each sample was ground, pressed into a pellet, and heated in a final step of $1200^{\circ}$C for 20 hours. After this procedure, dark-black materials were obtained in 10 mm diameter disks. High-resolution XRD data were collected for each sample at room temperature using a PANalytical Empyrean diffractometer, operating with Cu $K_{\alpha}$ radiation. The XRD data was investigated over the angular range $10^{\circ}\leq 2\theta\leq100^{\circ}$, with a 2${\theta}$ step size of 0.0065$^{\circ}$. Rietveld refinement was performed with GSAS software, and its graphical interface program \cite{GSAS}. DC M(T) and M(H) measurements were carried out using a Quantum Design PPMS-VSM magnetometer. The M(T) curves were measured in both zero field cooled (ZFC), and field cooled (FC) modes. For the ZFC M(H) measurement, each sample was demagnetized with an oscillating field at room temperature prior to the experiment to eliminate any small trapped field in the magnet.

\section{RESULTS AND DISCUSSION}

\begin{figure}
\begin{center}
\includegraphics[width=1.2 \textwidth]{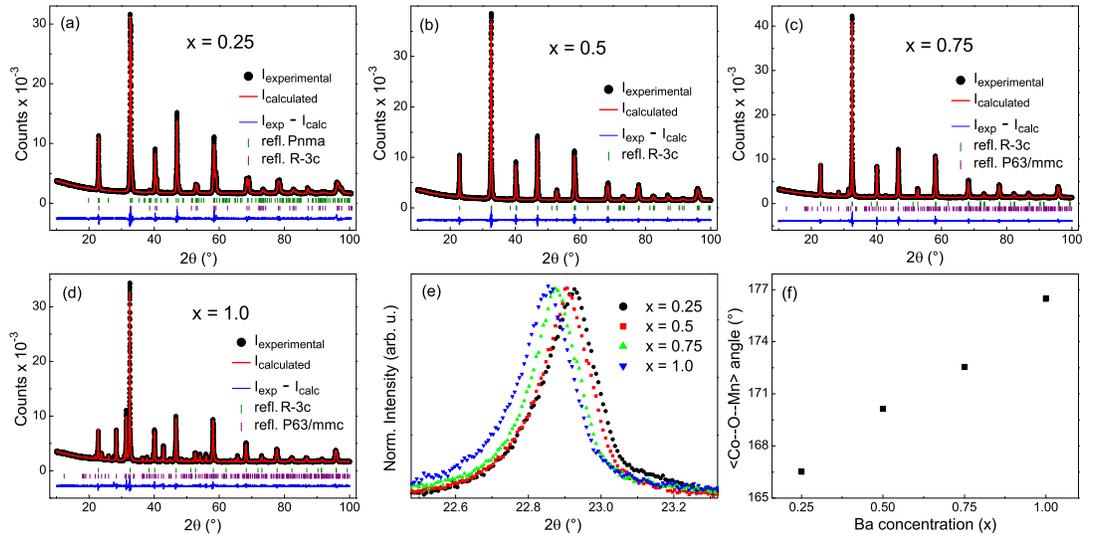}
\end{center}
\caption{(a)-(d) Rietveld refinement fittings of La$_{2-x}$Ba$_{x}$CoMnO$_{6}$ samples. The solid black circles represent the experimental data. The solid red lines are for the calculated patterns. The blue line shows the difference between the observed and calculated patterns. Vertical lines represent the Bragg reflections for the respective space groups. (e) Normalized first Bragg reflection of each sample, evidencing the shift toward left when the Ba-content increases. (f) Evolution of the pseudo linear $\langle$Co--O--Mn$\rangle$ bond angle as a function of Ba-concentration, $x$.}
\label{Fig_XRD}
\end{figure}

The XRD patterns of La$_{2-x}$Ba$_{x}$CoMnO$_{6}$ samples here investigated (hereafter called by its Ba-concentration, $x$) are shown in Figs. \ref{Fig_XRD}(a)-(d), while the main results obtained from the Rietveld refinement are displayed in Table \ref{T1}. The refinement indicates that $x$ = 0.25 compound crystallizes with two crystallographic phases, with $\sim20\%$ of its mass weight in the orthorhombic $Pnma$ space group and $\sim80\%$ in the rhombohedral $R\bar{3}c$ one. This is reasonable, since $x$ = 0 parent compound forms in $Pnma$ space group \cite{PRB2019} while the $x$ = 0.5 sample belongs to $R\bar{3}c$ \cite{BJP2020}. As the Ba-concentration increases, a tendency toward hexagonal symmetry is observed, with $x$ = 0.75 and 1.0 samples forming with $\sim5\%$ and $\sim30\%$ of $P6_{3}/mmc$ space group, respectively. These results agree with previous works reporting this hexagonal space group for Ba-rich perovskites \cite{Miranda}. However, it is important to note that even the highest Ba-doping level investigated here is insufficient to ensure a complete structural transition to this space group. The $x$ = 1.0 sample still keeps $\sim70\%$ of the $R\bar{3}c$ rhombohedral symmetry.

Fig. \ref{Fig_XRD}(e) compares the first Bragg reflection of the samples here investigated, showing a clear tendency towards the left along the 2$\theta$ axis,  suggesting an expansion of the unit cell due to the Ba-doping. This drift confirms the incorporation of this ion in the lattice since in XII coordination, the Ba$^{2+}$ ionic radius (1.61 \AA) is significantly larger than that of La$^{3+}$ (1.36 \AA) \cite{Shannon}. Corroborating also the tendency of higher symmetry space group with the Ba to La substitution since the crystal symmetry is generally higher for DPs with larger ionic radius at A-site \cite{Serrate,Sami}. Such a trend is manifested in the average Co--O--Mn bond angle that increases with Ba-doping, as depicted in Fig. \ref{Fig_XRD}(f). This structural change directly impacts the orbital hybridization between the TM ions and affects the system's magnetic properties.

\begin{table}
\centering
\caption{Results of the Rietveld refinement of crystalline structures for La$_{2-x}$Ba$_{x}$CoMnO$_{6}$ samples.}
\label{T1}
\resizebox{\textwidth}{!}{\begin{tabular}{c|ccccccc}
\hline
\hline
Sample & \multicolumn{2}{c}{x = 0.25} & 0.5 & \multicolumn{2}{c}{x = 0.75} & \multicolumn{2}{c}{x = 1.0} \\
\hline
Space group & $Pnma$ & $R\bar{3}c$ & $R\bar{3}c$ & $R\bar{3}c$ & $P6_{3}/mmc$ & $R\bar{3}c$ & $P6_{3}/mmc$ \\

Mass weight (\%) & 19.3 & 80.7 & 100 & 95.1 & 4.9 & 72.7 & 27.3 \\
\hline
$a$ (\AA) & 5.3721(5) & 5.4755(1) & 5.4908(1) & 5.4925(1) & 5.6826(3) & 5.4912(6) & 5.6819(1) \\
$b$ (\AA) & 7.7604(11) & 5.4755(1) & 5.4908(1) & 5.4925(1) & 5.6826(3) & 5.4912(6) & 5.6819(1 \\
$c$ (\AA) & 5.4599(8) & 13.2556(2) & 13.3676(2) & 13.4281(5) & 28.5099(20) & 13.4470(5) & 28.4792(8) \\
\hline

$\langle$Co/Mn--O$\rangle$ (\AA) & \multicolumn{2}{c}{1.8957(2)} & 1.9445(1)  & \multicolumn{2}{c}{1.9416(1)} & \multicolumn{2}{c}{1.9278(1)} \\

$\langle$Co--O--Mn$\rangle$ ($^{\circ}$) & \multicolumn{2}{c}{166.52(1)} & 170.14(1) & \multicolumn{2}{c}{172.55(1)} & \multicolumn{2}{c}{176.47(1)} \\

\hline
$R_{wp}$ ($\%$) & \multicolumn{2}{c}{3.6} & 3.5  & \multicolumn{2}{c}{3.6} & \multicolumn{2}{c}{3.6} \\
$R_{p}$ ($\%$) & \multicolumn{2}{c}{2.7} & 2.6  & \multicolumn{2}{c}{2.7} & \multicolumn{2}{c}{2.7} \\
\hline
\hline
\end{tabular}}
\end{table}

\begin{figure}
\begin{center}
\includegraphics[width=0.7 \textwidth]{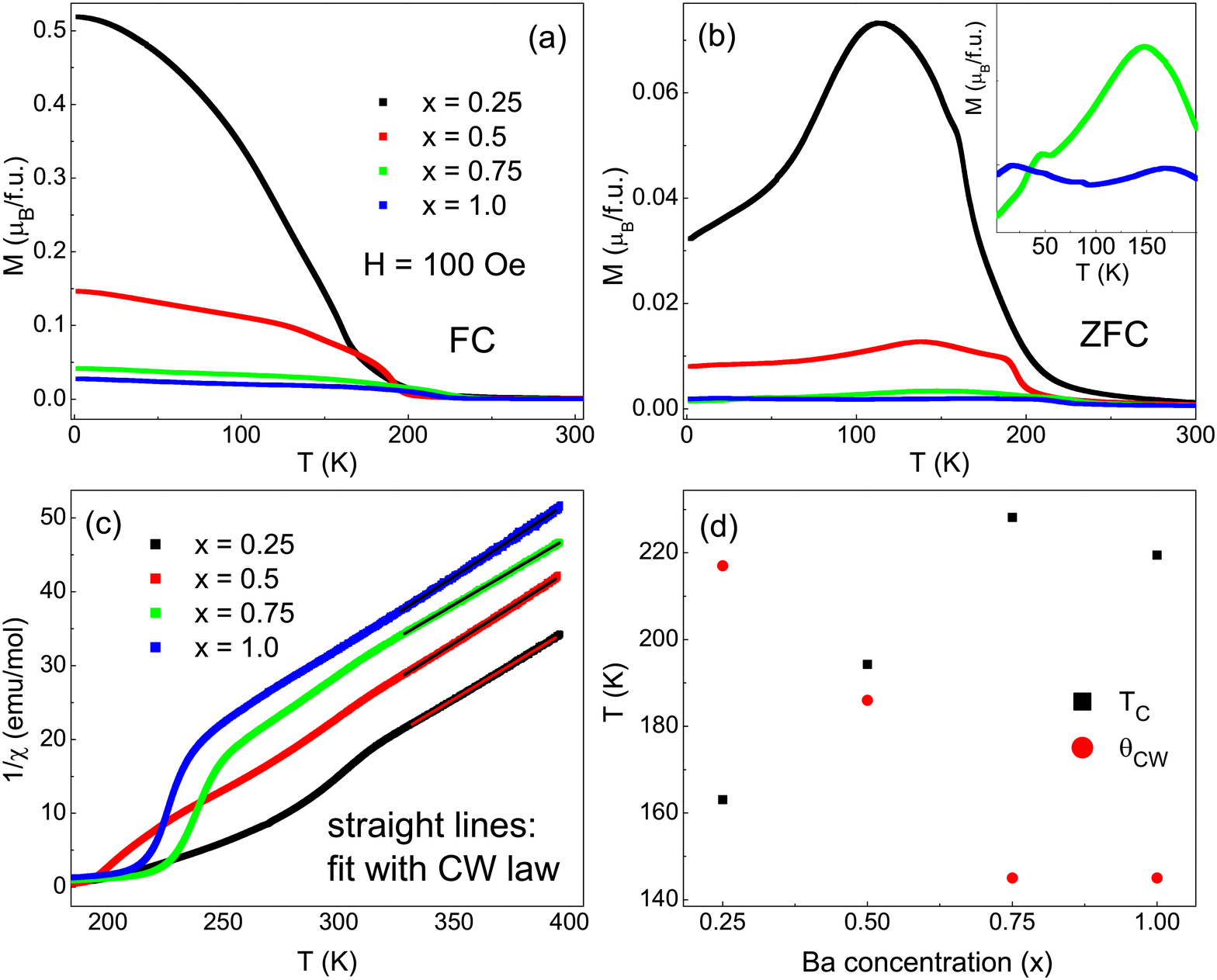}
\end{center}
\caption{(a) FC and (b) ZFC $M(T)$ curves of La$_{2-x}$Ba$_{x}$CoMnO$_{6}$ samples, measured with $H$ = 100 Oe. The inset shows a magnified view of the low temperature region of the $x$ = 0.75 and 1.0 curves. (c) $\chi^{-1}$ as a function of temperature, where the straight lines represent the best fits of the PM region with CW law. (d) Evolution of $T_C$ and $\theta_{CW}$ with Ba-doping.}
\label{Fig_MxT}
\end{figure}

Fig. \ref{Fig_MxT}(a) displays the FC M(T) curves of the investigated samples. All curves show FM-like behavior, usually ascribed to Co$^{2+}$--O--Mn$^{4+}$ and Co$^{3+}$--O--Mn$^{3+}$ positive exchange couplings \cite{PRB2019}. However, the ASD promotes some short-range AFM interactions such as Co$^{3+}$--O--Co$^{3+}$, Mn$^{4+}$--O--Mn$^{4+}$ and other couplings, which are evidenced by the ZFC curves displayed in Fig. \ref{Fig_MxT}(b), where the inset shows some additional anomalies in the low-$T$ region of the $x$ = 0.75 and 1.0 curves. The Curie-Weiss (CW) temperature ($\theta_{CW}$), obtained from the fit of the paramagnetic (PM) region of the magnetic susceptibility curve with the CW law [Fig. \ref{Fig_MxT}(c)], is displayed in Fig. \ref{Fig_MxT}(d) and Table \ref{T2}. Its decrease with Ba-doping may be related to the increased portion of AFM phases, although the positive values confirm that FM interactions are dominant for all compounds.

Interestingly, the magnetic transition temperature ($T_C$) increases with Ba-doping up to $x$ = 0.75, followed by a decrease for $x$ = 1.0 [Fig. \ref{Fig_MxT}(d)]. This trend may result from a joint action of structural and electronic changes. As previously discussed, the Ba$^{2+}$ substitution at La$^{3+}$ site leads to the increase of the Co--O--Mn bond angle, favoring the Co $e_g$--Mn $e_g$ orbital hybridization and thus promoting the increase of $T_C$. On the other hand, the high Ba$^{2+}$-doping level for $x$ = 1.0 sample signifies a large amount of Co$^{3+}$ and a small amount of Mn$^{3+}$ that reduce the percolation of Co$^{2+}$--O--Mn$^{4+}$ and Co$^{3+}$--O--Mn$^{3+}$ FM couplings, thus decreasing $T_C$. The non-monotonic evolution of the effective magnetic moment ($\mu_{eff}$) obtained from the CW-fit (see Table \ref{T2}) is also a manifestation of the delicate correlation between the crystal and electronic structures in this system, which may affect not only the Co/Mn valences but also their orbital moments and spin states. The disorder and competing magnetic phases that result from such complex balance are key ingredients for the appearance of SG-like phases, as already reported for $x$ = 0.5 \cite{BJP2020}. Glassy magnetism is, in turn, ubiquitous in SEB compounds \cite{Model,Model2,PRM2021}.

\begin{table}
\centering
\caption{Main parameters obtained from the $M(T)$ and $M(H)$ measurements.}
\begin{tabular}{c|cccc}
\hline
\hline
Sample & 0.25 & 0.5 & 0.75 & 1.0 \\
\hline
$T_C$ (K) &  163 & 194 & 228 & 219 \\
$\theta_{CW}$ (K) & 217$\pm$1 & 186$\pm$1 & 145$\pm$1 & 145$\pm$2 \\
$\mu_{eff}$ ($\mu_B$/f.u.) & 6.46$\pm$0.01 & 6.31$\pm$0.01 & 6.56$\pm$0.01 & 6.25$\pm$0.01 \\
\hline
$H_{EB}$ (Oe) & 666 & 2657 & 1872 & 1562 \\
$H_{C}$ (kOe) & 7.53 & 4.00 & 3.68 & 2.21 \\
$M_R$ ($\mu_B$/f.u.) & 1.393 & 0.117 & 0.052 & 0.028 \\
\hline
\hline
\end{tabular}
\label{T2}
\end{table}

\begin{figure}
\begin{center}
\includegraphics[width=0.7 \textwidth]{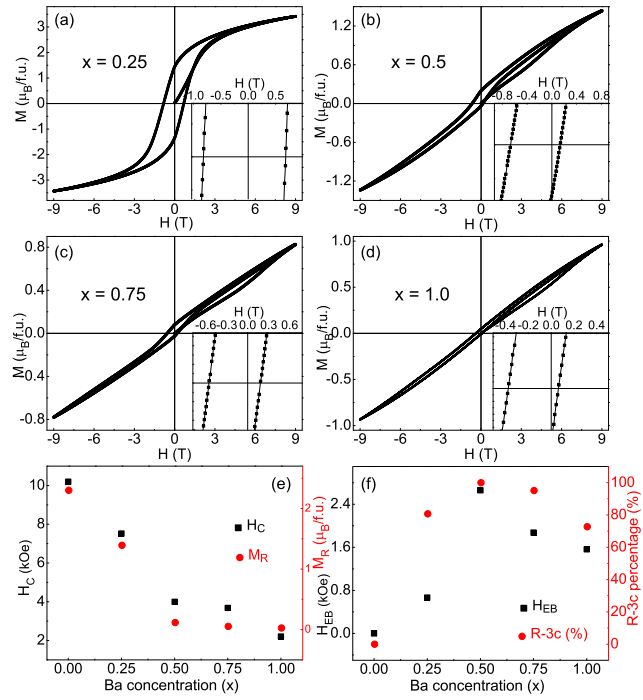}
\end{center}
\caption{(a)-(d) $M(H)$ loops for La$_{2-x}$Ba$_{x}$CoMnO$_{6}$ measured at 5 K after ZFC each sample. Insets show magnified views of the regions near $M$ = 0, highlighting the horizontal shifts along the $H$ axis. (e) $M_R$ and $H_C$ as a function of Ba-concentration, $x$. (f) $H_{EB}$ and portion of $R\bar{3}c$ space group as a function of $x$, where the values for $x$ = 0 were extracted from Ref. \cite{LaCa2022}.}
\label{Fig_MxH}
\end{figure}

Figs. \ref{Fig_MxH}(a)-(d) show the M(H) curves measured after ZFC each sample down to 5 K. All loops are closed and show some hysteresis. However, they gradually lose their squareness with increasing $x$ and develop a roughly linear $H$-dependence at high fields, further indicating the increase in the proportion of AFM interactions with $x$. Such increase is corroborated by the systematic decrease of the remanent magnetization ($M_R$) and coercive field [$H_{C}=(H_{C+}-H_{C-})/2$, where $H_{C+}$ and $H_{C-}$ are respectively the positive and negative coercive fields], as depicted in Fig. \ref{Fig_MxH}(e).

Notably, all M(H) curves are shifted toward the left along the $H$ axis, characterizing the SEB effect. As can be seen in Fig. \ref{Fig_MxH}(f) and Table \ref{T2}, the EB field, $H_{EB}=|H_{C+}+H_{C-}|/2$, is of the same order of magnitude of the largest SEB values reported so far \cite{Giri,Zhang,Murthy}. It can also be noticed in Figs. \ref{Fig_MxH}(a)-(d) that some stretch of the virgin curves lies outside the main hysteresis loops. This characteristic feature of SEB materials is usually related to $H$-induced reconfiguration of the spins at the magnetic interfaces that occur during the initial magnetization of the system \cite{Wang,Murthy,JMMM2017}.

Fig. \ref{Fig_MxH}(f) shows that the $H_{EB}$ is maxima for $x$ = 0.5, in resemblance with La$_{2-x}$Sr$_x$CoMnO$_6$ and La$_{2-x}$Ca$_x$CoMnO$_6$ series growth by the sol-gel method \cite{Murthy2, Sahoo3}. The evolution of $H_{EB}$ for these Sr- and Ca-based systems was ascribed to changes in the ASD at Co/Mn site, which destroys the long-range magnetic order and gives rise to a metamagnetic state at low temperatures. Contrastingly, for La$_{2-x}$Ca$_x$CoMnO$_6$ samples growth by solid-state reaction, it was observed a systematic increase of $H_{EB}$ with $x$, which was explained in terms of the increasing portion of the AFM phases present in the system \cite{LaCa2022}.

For the Ba-based samples investigated here, the changes in $H_{EB}$ may be related to the structural evolution of the system. Fig. \ref{Fig_MxH}(f) shows the percentage of rhombohedral $R\bar{3}c$ space extracted from the Rietveld refinements. For the $x$ = 0 parent compound there is no $R\bar{3}c$ phase, and no $H_{EB}$ \cite{LaCa2022}. This rhombohedral phase appears with Ba-doping and increases up to $x$ = 0.5, where 100\% of the crystal belongs to $R\bar{3}c$ space group and $H_{EB}$ achieves its maximum value. Then, for $x$ $>$ 0.5, the hexagonal $P6_3/mmc$ emerges and increases with increasing the Ba-content, leading to the concomitant decrease of the $R\bar{3}c$ portion, where the $H_{EB}$ decreases. It remains unclear in this work how precisely the crystal structure affects the $H_{EB}$. It could be related, for instance, to ASD since crystallographic symmetry is known to impact the permutation between the TM ions in the lattice. However, the $R\bar{3}c$ space group is viewed as a disordered system for which the Co and Mn ions share the same site, precluding the estimation of the ASD from our XRD data. A detailed investigation of the crystal structure using anomalous scattering in XRD, Raman spectroscopy, neutron diffraction, and other techniques would be necessary to unravel this issue.

\section{CONCLUSIONS}

In summary, the La$_{2-x}$Ba$_{x}$CoMnO$_{6}$ ($0.25 \leq x \leq 1.0$) polycrystalline samples were successfully synthesized by solid-state reaction. The crystal structure evolves from orthorhombic $Pnma$ to rhombohedral $R\bar{3}c$ and then to hexagonal $P6_{3}/mmc$ space group as the Ba-concentration increases, but even the largest Ba-doping here investigated is not enough to ensure a complete transition to the hexagonal symmetry, with part of the system remaining in the $R\bar{3}c$ space group. Although an FM-like behavior is observed for all samples, an AFM phase seems to be present and increases with Ba-doping. The competing FM and AFM interactions give rise to the SEB phenomena. The evolution of $H_{EB}$ follows the same trend of the percentage of $R\bar{3}c$ phase along the series, suggesting that the crystal structure plays a major role in this effect for La$_{2-x}$Ba$_{x}$CoMnO$_{6}$.

\section{ACKNOWLEDGMENTS}
This work was supported by Conselho Nacional de Desenvolvimento Cient\'{i}fico e Tecnol\'{o}gico (CNPq) [No. 425936/2016-3], Coordena\c{c}\~{a}o de Aperfei\c{c}oamento de Pessoal de N\'{i}vel Superior (CAPES), Funda\c{c}\~{a}o de Amparo \`{a} Pesquisa do Estado de Goi\'{a}s (FAPEG), and Funda\c{c}\~{a}o Carlos Chagas Filho de Amparo \`{a} Pesquisa do Estado do Rio de Janeiro (FAPERJ) [E-26/211.291/2021].

\end{document}